\documentclass[12pt]{article}
\usepackage{amsmath}
\usepackage{amsfonts} 
\usepackage{graphicx,psfrag,epsf}
\usepackage{enumerate}
\usepackage{natbib}
\usepackage{color}
\usepackage{url} 
\usepackage{tabularx} 
\usepackage{multirow} 
\usepackage{stackrel} 
\usepackage[font={small}]{caption} 

\newcommand{\blind}{1}

\begin{document}

\def\spacingset#1{\renewcommand{\baselinestretch}%
{#1}\small\normalsize} \spacingset{1}


\if1\blind
{
  \title{\vspace*{-35pt} \bf Diagnosing Glaucoma Progression with Visual Field Data Using a Spatiotemporal Boundary Detection Method} 
  \date{}
  \author{Samuel I.\ Berchuck\thanks{Samuel I.\ Berchuck is a Postdoctoral Associate, Department of Statistical Science and Forge, Duke University, NC 27708 (E-mail: sib2@duke.edu). Jean-Claude Mwanza is an Assistant Professor, Department of Ophthalmology, University of North Carolina-Chapel Hill (DO, UNC-CH), NC 27517 (E-mail: jean-claude\_mwanza@med.unc.edu). Joshua L.\ Warren is an Assistant Professor, Department of Biostatistics, Yale University, New Haven, CT, 06520 (E-mail: joshua.warren@yale.edu). This work was partially supported by the National Institute of Environmental Health Sciences (SIB; T32ES007018), the Research to Prevent Blindness (DO, UNC-CH), and the National Center for Advancing Translational Science (JLW; UL1 TR001863, KL2 TR001862). The authors thank the Lions Eye Institute for Figure \ref{fig:figure1}b and Wallace L.M.\ Alward (Ophthalmology and Visual Sciences, University of Iowa Carver College of Medicine) for Figure \ref{fig:figure1}c; and also Brigid D.\ Betz-Stablein (School of Medical Sciences, University of New South Wales), William H.\ Morgan (Lions Eye Institute, University of Western Australia), Philip H.\ House (Lions Eye Institute, University of Western Australia), and Martin L.\ Hazelton (Institute of Fundamental Sciences, Massey University) for providing the dataset from their original study for use in this analysis.
}\hspace{.2cm}, Jean-Claude Mwanza, and Joshua L.\ Warren}
  \maketitle
} \fi

\if0\blind
{
  \bigskip
  \bigskip
  \bigskip
  \begin{center}
    {\LARGE \vspace*{75pt} \bf Diagnosing Glaucoma Progression with Visual Field Data Using a Spatiotemporal Boundary Detection Method}
\end{center}
  \medskip
} \fi

\vspace*{-20pt}

\begin{abstract}
Diagnosing glaucoma progression is critical for limiting irreversible vision loss. A common method for assessing glaucoma progression uses a longitudinal series of visual fields (VF) acquired at regular intervals. VF data are characterized by a complex spatiotemporal structure due to the data generating process and ocular anatomy. Thus, advanced statistical methods are needed to make clinical determinations regarding progression status. We introduce a spatiotemporal boundary detection model that allows the underlying anatomy of the optic disc to dictate the spatial structure of the VF data across time. We show that our new method provides novel insight into vision loss that improves diagnosis of glaucoma progression using data from the Vein Pulsation Study Trial in Glaucoma and the Lions Eye Institute trial registry. Simulations are presented, showing the proposed methodology is preferred over existing spatial methods for VF data. Supplementary materials for this article are available online and the method is implemented in the R package \texttt{womblR}. 
\end{abstract}

\noindent%

\newpage
\spacingset{1.45} 


\section{INTRODUCTION}
\label{sec:intro}

Glaucoma is a leading cause of blindness worldwide, with a prevalence of 4\% in the population aged 40-80 \citep{tham2014global}. The most common form of glaucoma is primary open-angle glaucoma (POAG). The biological basis of the disease is not fully defined, however the most significant risk factor for POAG is elevated intraocular pressure (IOP). Elevated IOP can be treated with eye drops, surgery, or laser. If this condition goes untreated the optic nerve may be damaged, resulting in permanent vision loss. Since visual impairment caused by glaucoma is irreversible and efficient treatments exist, early detection of the disease is essential. As such, patients diagnosed with glaucoma are monitored for disease progression even if they are receiving treatment, because the role of the treatment is to slow the progression. Determining if the disease is progressing remains one of the most challenging aspects of glaucoma management, since it is difficult to distinguish true progression from variability due to natural degradation or noise \citep{vianna2015detect}. Numerous techniques have been developed to monitor progression, but there is currently no consensus as to which method is best. In this study we focus on visual field (VF) testing.

A VF test is a psychophysical procedure that assesses a patient's field of vision. The test results in a 2-dimensional map that represents the level of eyesight uniformly across the retina. Glaucoma patients normally receive bi-annual VF tests and have follow-up lasting numerous years \citep{chauhan2008practical}. The collection of VF data results in a longitudinal series of spatially referenced measurements that exhibit a complex spatiotemporal structure. The spatial surface of the VF is observed on a lattice (i.e., uniform areal data), however it is a complex projection of the underlying optic disc and exhibits anatomically induced spatial dependencies. In particular, localized damage to the optic disc can result in clinically deterministic deterioration across the VF \citep{quigley1992evaluation}. Incorporating this non-standard spatial dependence structure into our methodology is a priority for properly analyzing these data. 

There are comparable methodological complexities that arise in Alzheimer's disease, attention deficit hyperactivity disorder, and multiple sclerosis, all related to white-matter connectivity of the brain \citep{he2008structural, konrad2010adhd}. However, complex brain imaging data are generally analyzed using point-referenced statistical models as opposed to areal data models \citep{bowman2008bayesian}. The point-referenced framework has a rich theory that accounts for non-standard spatial dependencies, mainly through the assumptions of non-stationarity \citep{sampson2010constructions} and anisotropy \citep{ecker2003spatial}. \cite{castruccio2016multi} model anatomical regions of interest of the brain in fMRI data using these assumptions in a study of stroke rehabilitation, while another study accounts for brain connectivity in Alzheimer's patients \citep{thompson2004mapping}.

The literature surrounding complex spatial dependencies is less developed in the areal data setting with spatiotemporal methods even less common. One reason for this is that stationarity and isotropy (i.e., correlation as a function of distance alone) cannot be defined for areal data due to the contrasting definition of spatial proximity between the two frameworks. When analyzing areal data, spatial similarity is typically dictated by a local neighborhood structure \citep{banerjee2003hierarchical}. Over time, modifications to the neighborhood structure have been proposed, and in this manuscript we work within the boundary detection framework. An extension of directional gradients from point-referenced theory \citep{banerjee2006bayesian}, boundary detection was originally developed to identify boundaries on geographical maps in the context of disease mapping \citep{ma2007bayesian}. The inherited method of modifying the local neighborhood structure provides an intricate framework for introducing complex spatial structure on the VF. We introduce a novel spatiotemporal boundary detection model that allows the underlying anatomy of the optic disc to dictate the spatial structure of the VF across time and show that it offers new and valuable information for improved progression detection through analysis of data from the Vein Pulsation Study Trial in Glaucoma (VPSG) and the Lions Eye Institute trial registry.

This paper is outlined as follows. Section \ref{sec:vf} details the data generating mechanism for VF data and the setting of glaucoma progression diagnostics. We briefly review spatial boundary detection methods in Section \ref{sec:bd}. In Section \ref{sec:methods}, our newly developed statistical methodology is described. We apply our method to a dataset of VF tests from glaucoma patients in Section \ref{sec:da} and compare its performance to an existing boundary detection method via simulation study in Section \ref{sec:sim}. We conclude in Section \ref{sec:disc} with a discussion.


\section{VISUAL FIELD DATA}
\label{sec:vf}

\begin{figure}
\begin{center}
\includegraphics[scale=0.69]{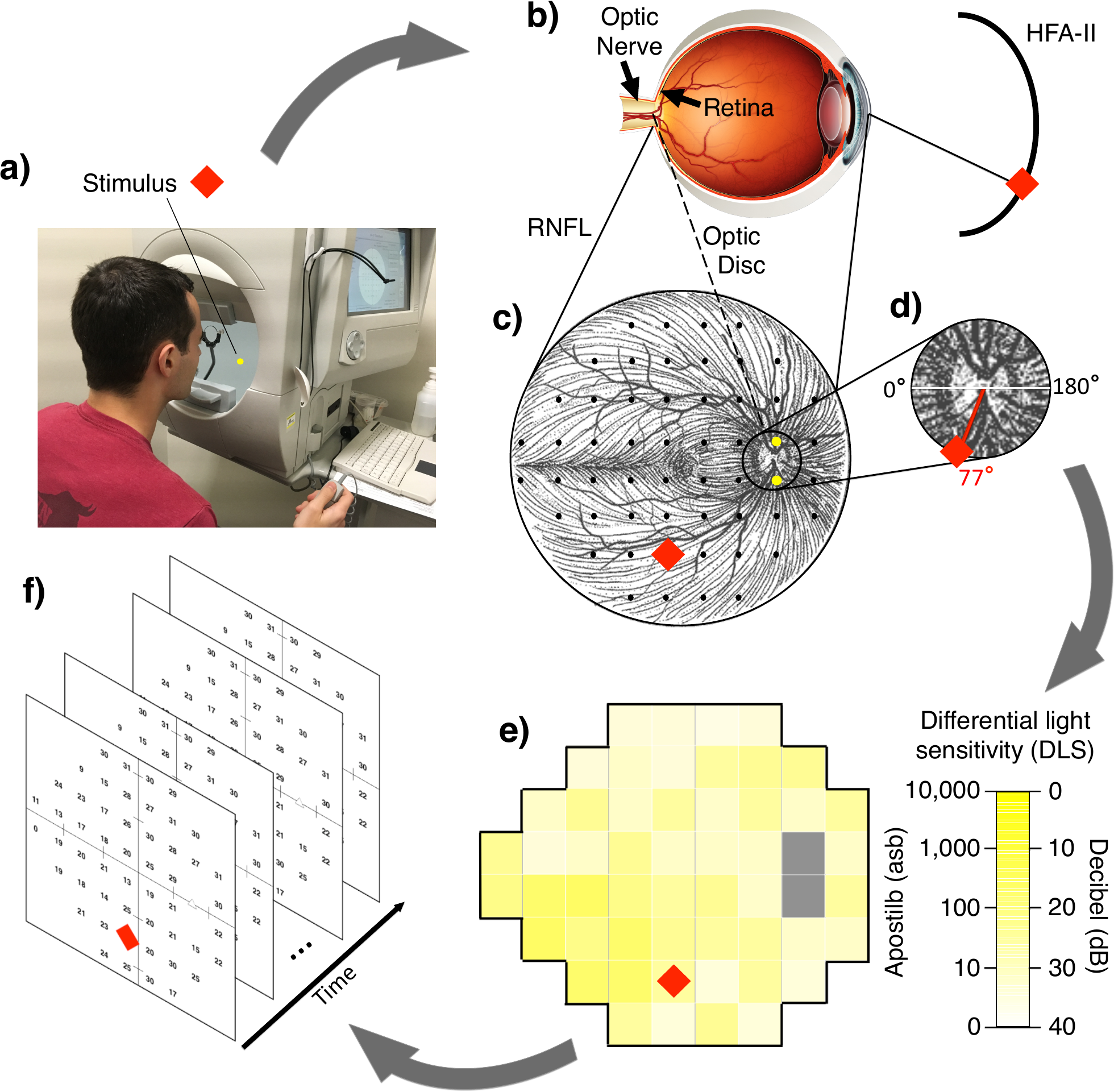}\\
\caption{Demonstrating the VF data generating mechanism, \textbf{a)} A patient (the first author) participates in a VF examination. \textbf{b)} The HFA-II stimulus is absorbed by the retina and transmitted through the optic disc, and along the optic nerve to the brain. \textbf{c)} Each VF test point corresponds to a location on the underlying RNFL. \textbf{d)} The corresponding nerve fiber enters the optic disc at 77$^\circ$. \textbf{e)} Each VF test produces a map that shows the intensity a stimulus is detected at each test location. The gray locations represent the blind spot. \textbf{f)} Over time the patient accrues VF tests.\label{fig:figure1}}
\end{center}
\end{figure}

The VF is the spatial array of visual sensations that the brain perceives as vision \citep{smythies1996note}. The most common technique for testing the VF is standard automated perimetry (SAP) \citep{chauhan2008practical}. In this study we analyze data acquired with the Humphrey Field Analyzer-II (HFA-II) (Carl Zeiss Meditec Inc., Dublin, CA). The VF data generating process is displayed in Figure \ref{fig:figure1}. We follow a single observation throughout the figure, presented as a diamond. In Figure \ref{fig:figure1}a, a patient (the first author) is tested on a HFA-II. SAP constructs a VF map by assessing a patient's response to varying levels of light. Patients are instructed to focus on a central fixation point as light is introduced randomly in a preceding manner over a grid on the VF. As light is observed, the patient presses a button and the current light intensity is recorded \citep{chauhan2008practical}. The process is repeated until the entire VF is tested. In the figure, the first author stares at the background of the machine, waiting until he observes the stimulus to press the buzzer.

The HFA-II measures 54 test points, however, two of these correspond to a natural blind spot corresponding to the optic disc, resulting in 52 informative points. The points that correspond to the blind spot are highlighted yellow (Figure \ref{fig:figure1}c). Figure \ref{fig:figure1}b presents the anatomy of the eye. The retina is a light-sensitive layer at the back of the eye that absorbs stimulus from the HFA-II and transmits information to the brain through the optic disc along the optic nerve. The optic nerve is a bundle of more than one million fibers that carry visual information in the form of electrical signal from the retina to the brain. The retinal nerve fiber layer (RNFL), Figure \ref{fig:figure1}c, is made of retinal ganglion cell (RGC) axons. All the axons converge at the optic disc, which is the departure point of the optic nerve. 

These RGC axons are responsible for encoding visual information. The ganglion cells disperse across the retina, but are mostly concentrated in the center of the retina, and use photoreceptors to transmit information to the brain along their axons \citep{davson2012physiology}. Both the RGCs and their axons may die progressively as a result of elevated IOP. In particular, damage to specific regions of the optic disc corresponds to loss of RGCs whose axons enter the damaged region. Thus, vision loss across the VF and the corresponding damage to the optic disc are the result of the death of RGCs and their axons. Furthermore, correlation between two test points on the VF is dependent on the spatial proximity that their underlying nerve fibers enter the optic disc. This indicates that variability in the neighborhood structure of the VF is possibly indicative of progression. The nerve fibers (not to be confused with the thicker and more sparse blood vessels) are shown extending throughout the retina with the VF test points represented by black dots. 

Each test location has underlying fibers that track across the RNFL and enter the optic disc at a particular angle. \cite{garway2000mapping} studied the relationship between the VF test points and the underlying RNFL. They quantified the relationship between the VF and optic disc by estimating the angle that each test location's underlying fiber enters the optic disc, measured in degrees ($^\circ$). The measure ranges from 0-360$^\circ$, where 0$^\circ$ is designated at the 9-o'clock position (right eye) and angles are counted counter clockwise. The Garway-Heath angle for the example location is 77$^\circ$ (Figure \ref{fig:figure1}d).

In the course of a VF test all 52 informative locations are assessed, resulting in a grid of test values (Figure \ref{fig:figure1}e). The two locations corresponding to the blind spot are gray. SAP measures the differential light sensitivity (DLS) across the VF. The measurement represents a contrast between the background of the machine, normally white, and the light stimulus. The intensity of the stimulus is initially similar to the background, but as the intensity increases the contrast grows and the probability of detecting the stimulus increases. The intensity of the stimulus is measured in Apostilbs (asb), where larger values represent a greater intensity. The HFA-II is capable of outputting intensities ranging from one (similar to the background) to 10,000 asb. All stimuli that are not detected by 10,000 asb are censored, due to the constraints of the machine. In practice, these intensities are converted to decibels (dB), where dB $=40-10\log_{10}$ (asb). This inverts the scale, such that DLS values near 40 indicate good vision, while values near zero represent possible blindness. In the course of monitoring a patient with glaucoma, VF testing is performed on a regular basis and a longitudinal series of VFs is obtained (Figure \ref{fig:figure1}f). 


\section{SPATIAL BOUNDARY DETECTION}
\label{sec:bd}

VF data exhibit a complex spatiotemporal structure that is characterized by localized spatial dependencies dictated by the underlying anatomy of the optic disc. These data are generated over a lattice, a subset of areal data in spatial literature. In spatial statistics, the foundational assumption states that dependence between observations weakens as the distance between locations increases \citep{gelfand2010handbook}. In areal data models, this assumption often manifests through Gaussian Markov random fields (GMRF) that induce neighborhood dependence across the region \citep{geman1984stochastic}. A common GMRF is the conditional autoregressive (CAR) process \citep{besag1974spatial}. 

The CAR model achieves spatial smoothing through random effects, $\varphi_i$ at location $i$ for $i=1,\ldots,n$, with spatial structure defined through the set of neighborhood adjacencies, $\{w_{ij}\}$. These adjacencies are fixed, such that $w_{ij}=1(i \sim j)$, where $1(\cdot)$ is the indicator function and $i \sim j$ is the event that locations $i$ and $j$ share a border ($w_{ii}=0$ for all $i$). This specification induces an elegant conditional distribution for each random effect. We present a general form of the CAR process originally introduced by \cite{leroux2000estimation}, 
\begin{equation} \label{eq:car}
\varphi_i|\boldsymbol{\varphi}_{-i},\mu,\tau^2 \sim \text{N}\left(\frac{\rho\sum_{j=1}^{n}w_{ij}\varphi_j + (1-\rho)\mu}{\rho\sum_{j=1}^{n}w_{ij} + 1-\rho}\hspace{0.2cm},\hspace{0.2cm} \frac{\tau^2}{\rho\sum_{j=1}^{n}w_{ij} + 1-\rho}\right),
\end{equation}
where $\boldsymbol{\varphi}_{-i}=(\varphi_1,\ldots,\varphi_{i-1},\varphi_{i+1},\ldots,\varphi_n)^T$. Note that setting $\rho=1$ reduces Equation \ref{eq:car} to the standard intrinsic CAR process. The mean is a weighted average of the neighbors with variance decreasing inversely with the number of neighbors. The standard CAR model provides an attractive representation and is flexible in its ability to model smooth spatial processes. However, it can be limited in settings where spatial structure is fragmented into localized regions due to \{$w_{ij}$\} being fixed. In the areal data setting, a flexible class of models called boundary detection can be used to remedy this issue \citep{banerjee2003hierarchical}. 

Boundary detection was originally explored by \cite{womble1951differential}, but has gained a niche in the context of disease mapping. The motivation for boundary detection is to identify regions on the spatial surface where there are sharp changes in the response value \citep{jacquez2003international}.  In disease mapping, these boundaries can take many forms, for example geographic obstacles such as mountain ranges or socioeconomic boundaries caused by pockets of increased poverty. Standard methods attempt to control for disjoint spatial regions by including covariates in the mean structure of the CAR process. This technique can be effective in producing highly variable spatial surfaces, but is limited in producing truly localized spatial smoothing \citep{banerjee2003hierarchical}. Boundary detection improves on this naive approach by carefully considering the form of the adjacencies. 

Initially boundary detection methods were parameterized for use in disease mapping, defining boundaries as a function of the difference in standardized incidence ratios \citep{lu2005bayesian}. This method is limited, since it can be difficult to have knowledge of boundaries a priori and does not generalize outside of disease mapping. Numerous methods in the boundary detection literature treat the adjacencies as random variables and construct hierarchical models to estimate the adjacency matrix \citep{lu2007bayesian,ma2007bayesian}, and even provide extensions to the spatiotemporal setting \citep{rushworth2017adaptive}. However, inference from these models can be highly sensitive to prior specifications on certain parameters \citep{li2015bayesian}. Furthermore, these methods introduce $(n)(n-1)/2$ additional random variables leading to potential identifiability issues. \cite{li2011mining} proposed another class of methods that enumerate all possible permutations of the adjacencies in parallel models, using the Bayesian information criterion to choose between them. This class of methods was formalized when \cite{lee2014bayesian} introduced a novel joint prior distribution for the spatial random effects and adjacency matrix. This prior has been extended to spatiotemporal models \citep{lee2014controlling}.

A final class of methods models the adjacencies using dissimilarity metrics. The method introduced in \cite{lee2011boundary} generalizes the form of Equation \ref{eq:car} to allow for the adjacency weights to be modeled as a function of a small number of regression parameters, $\boldsymbol{\alpha}=(\alpha_1,\ldots,\alpha_q)^T$. According to \cite{lee2011boundary}, ``boundaries in the risk surface are likely to occur between populations that are very different because homogeneous populations should have similar risk profiles". They define $q$ non-negative dissimilarity metrics $\boldsymbol{z}_{ij}=(z_{ij1},\ldots,z_{ijq})^T$, where $z_{ijk}=|z_{ik}-z_{jk}|$ for $k=1,\ldots,q$. The $q$ covariates, $z_{ik}$ at location $i$, drive detection of boundaries and are characterized by their importance in defining the neighborhood structure. The choice of $q$ is problem specific and based on the availability of useful explanatory information for describing the boundaries. The adjacencies are modeled as follows, 
\begin{equation} \label{eq:duncanweights}
w_{ij}\left(\boldsymbol{\alpha}\right)= 1(i\sim j) 1\left(\exp\left\{- \boldsymbol{z}_{ij}^T \boldsymbol{\alpha}\right\} \geq 0.5\right)
\end{equation}
where each $\alpha_k$ is constrained to be non-negative so that a larger dissimilarity metric indicates a higher likelihood of a boundary (or zero weight). In this model, $\rho$ is fixed at 0.99 to force the spatial correlation structure to be determined locally by $\{w_{ij}\left(\boldsymbol{\alpha}\right)\}$, rather than globally by $\mu$. This form has many appealing properties that make it amenable to boundary detection. In particular, if there are no adjacencies (i.e., $w_{ij}(\boldsymbol{\alpha})=0$ for all $i\neq j$) the conditional mean and variance are still defined. The method proposed by \cite{lee2011boundary} provides a parsimonious framework for introducing localized smoothing.


\section{METHODS}
\label{sec:methods}

Following the approach of \cite{lee2011boundary}, we propose modeling localized spatial correlation through a set of weights $\{w_{ij}\left(\boldsymbol{\alpha}_t\right)\}$ as a parsimonious function of dissimilarity metrics and their regression parameters. However, we propose extending the framework to account for spatiotemporal localized smoothing and therefore define $\boldsymbol{\alpha}_t = (\alpha_{t1}, \ldots, \alpha_{tq})^T$, for $t = 1,\ldots,\nu$. With appropriate temporal dependency structures in place, this specification allows for localized smoothing in instances of true temporal correlation. Inference for this model is based on Markov chain Monte Carlo (MCMC) simulation, and a description of the algorithm is given in the online supplementary materials. Spatiotemporal models are computationally intensive, so the MCMC algorithm is implemented using Rcpp \citep{eddelbuettel2011rcpp} and is available from the R package \texttt{womblR} \citep{Rcore}. 


\subsection{Observational Model}
\label{sec:observational}

We begin by describing our new methodology generally before applying it to VF data in Section \ref{sec:da}. Let $Y_{it}$ denote an observation from spatial location $i$ at time $t$, $i=1,\ldots,n_t$, for $t=1,\ldots,\nu$. The number of locations can vary over time. Define $\boldsymbol{\varphi}_t=\left(\varphi_{1t},\ldots,\varphi_{n_t t}\right)^T$, with $\boldsymbol{\varphi}_{-it}$ missing the $i^\text{th}$ entry. The observational model is given by, 
\begingroup
\allowdisplaybreaks
\begin{align} \label{eq:observational}
Y_{it}|\vartheta_{it},\boldsymbol{\zeta} &\stackrel{\text{ind}}{\sim} f(Y_{it}|\vartheta_{it}, \boldsymbol{\zeta}) \quad \text{ for } i=1,\ldots,n_{t}, \text{ } t=1,\ldots,\nu,\\ \notag
g(\vartheta_{it})&=\varphi_{it},\\
\varphi_{it}|\boldsymbol{\varphi}_{-it},\mu_t,\tau_t^2,\boldsymbol{\alpha}_t &\stackrel{\text{ind}}{\sim} \text{N}\left(\frac{\rho\sum_{j=1}^{n_t}w_{ij}\left(\boldsymbol{\alpha}_t\right)\varphi_{jt} + (1-\rho)\mu_t}{\rho\sum_{j=1}^{n_t}w_{ij}\left(\boldsymbol{\alpha}_t\right) + 1-\rho}, \frac{\tau_t^2}{\rho\sum_{j=1}^{n_t}w_{ij}\left(\boldsymbol{\alpha}_t\right) + 1-\rho}\right). \notag
\end{align}
\endgroup

The parameter $\vartheta_{it}$ describes the distribution of $Y_{it}$ and our novel spatiotemporal random effect, $\varphi_{it}$, is introduced as a linear predictor of $g(\vartheta_{it})$, with $g(\cdot)$ a link function. Finally, $\boldsymbol{\zeta}$ is a vector of variance (or nuisance) parameters, for example the over-dispersion parameter in the negative binomial distribution. This modeling framework is general and accommodates generalized linear mixed models (GLMM). The GLMM setting can be obtained by setting $\vartheta_{it}=\mathbb{E}[Y_{it}|\vartheta_{it}]$. Due to the general specification, our methodology can be used to induce spatiotemporal localized smoothing in a general areal data setting, such as disease mapping.

The random effect for $\varphi_{it}$ represents an extension of the \cite{lee2011boundary} specification with temporally referenced parameters, $\mu_t$, $\tau_t^2$, and $\boldsymbol{\alpha}_t$ (referred to as observational level parameters). The $\rho$ parameter acts as a gauge of present spatial variation, where $\rho=0$ corresponds to global independence and $\rho \rightarrow 1$ implies strong spatial correlation. In the same vein as \cite{lee2011boundary}, we fix $\rho=0.99$ to guarantee that spatial correlation can be determined locally by the set of weights \{$w_{ij}(\boldsymbol{\alpha}_t)$\}.  We fully explore the impact of this decision through simulation as described in Section 5.6 and in the online supplementary materials.  Overall, we find that the results are robust to this choice.

The conditional distributions of the random effect can be written jointly using Brook's Lemma \citep{banerjee2003hierarchical},
$\boldsymbol{\varphi}_t|\mu_t,\tau_t^2,\boldsymbol{\alpha}_t \stackrel{\text{ind}}{\sim} \text{MVN}\left(\mu_t\boldsymbol{1}_{n_t}, \tau_t^2 \mathbf{Q}\left(\boldsymbol{\alpha}_t\right)^{-1}\right),$
$t=1,\ldots, \nu$, where $\mathbf{Q}\left(\boldsymbol{\alpha}_t\right)=\left[\rho \mathbf{W}^*\left(\boldsymbol{\alpha}_t\right) + (1-\rho)\mathbf{I}_{n_t}\right]$, and $\boldsymbol{1}_{n}$ and $\mathbf{I}_{n}$ are an $n$ dimensional column of ones and identity matrix, respectively. The matrix $\mathbf{W}^*\left(\boldsymbol{\alpha}_t\right)$ has diagonal elements $w^*_{ii}\left(\boldsymbol{\alpha}_t\right)=\sum_{j=1}^{n_t} w_{ij}\left(\boldsymbol{\alpha}_t\right)$ and off-diagonal elements $w^*_{ij}\left(\boldsymbol{\alpha}_t\right)=-w_{ij}\left(\boldsymbol{\alpha}_t\right)$.


\vspace{-0.01in}

\subsection{Neighborhood Model}
\label{sec:neighborhood}

We use a similar framework as \cite{lee2011boundary} to model the adjacency weights by writing them as a function of dissimilarity metrics, $\boldsymbol{z}_{ij}$. The weights are defined as follows, 

\begin{equation} \label{eq:weights}
w_{ij}\left(\boldsymbol{\alpha}_t\right)=1(i \sim j)\exp\left\{-\boldsymbol{z}_{ij}^T \boldsymbol{\alpha}_t\right\}.
\end{equation}
Unlike the original \cite{lee2011boundary} specification (Equation \ref{eq:duncanweights}), the weights are not forced to be binary. We do specify the components of $\boldsymbol{\alpha}_t$ to be non-negative, forcing the adjacencies in the open unit interval.  These constraints on $\alpha_{tk}$ yield intuitive interpretations at extreme values. As ${\alpha}_{tk} \rightarrow \infty$ the adjacencies become zero, resulting in an independence model, while ${\alpha}_{tk} \rightarrow 0$ reduces the adjacencies to a standard CAR process (Equation \ref{eq:car}). This new specification changes how neighbors share information. It is best understood through the conditional mean from Equation \ref{eq:car}, which becomes a weighted average of neighbors under our new definition of an adjacency, versus a simple average. 


\subsection{Temporal Model}
\label{sec:temporal}

From the joint specification of the random effects, we see that spatial structure is introduced through the covariance of $\boldsymbol{\varphi}_t$ at each time point and there is conditional independence ($\boldsymbol{\varphi}_{t_1} \perp \boldsymbol{\varphi}_{t_2}|\mu_t,\tau_t^2, \boldsymbol{\alpha}_t: t = t_1, t_2$). To induce temporal dependence between the $\boldsymbol{\varphi}_t$, we specify a separable temporal structure on the observational level parameters. Define,
\begin{equation} \notag 
\boldsymbol{\theta}=\left[\boldsymbol{\theta}_{\cdot 1} \cdots \boldsymbol{\theta}_{\cdot\nu}\right]=\left[ \begin{array}{ccc} \boldsymbol{\theta}_{1\cdot} \\ \boldsymbol{\theta}_{2\cdot} \\ \boldsymbol{\theta}_{2+1\cdot} \\ \vdots \\ \boldsymbol{\theta}_{2+q\cdot} \end{array} \right]=\left[ \begin{array}{ccc} \mu_1 & \cdots & \mu_{\nu}\\ \log\left(\tau_1\right) & \cdots & \log\left(\tau_{\nu}\right)\\ \log\left(\alpha_{11}\right) & \cdots & \log\left(\alpha_{\nu1}\right)\\ \vdots &  & \vdots \\ \log\left(\alpha_{1q}\right) & \cdots & \log\left(\alpha_{\nu q}\right)
\end{array} \right].
\end{equation}
Using properties of the vectorization function, $\text{vec}(\cdot)$, and the Kronecker product, $\otimes$, a separable process is specified such that
\begin{equation} \label{eq:separable}
\text{vec}\left(\boldsymbol{\theta}\right)|\boldsymbol{\delta},\mathbf{T},\phi \sim \text{MVN}\left(\boldsymbol{1}_{\nu} \otimes \boldsymbol{\delta}, \boldsymbol{\Sigma}\left(\phi\right) \otimes \mathbf{T}\right).
\end{equation}
This process yields elegant interpretations for the row and column moments of $\boldsymbol{\theta}$. From the moments, we see that $\boldsymbol{\delta}$ is a constant that corresponds to the mean of the observational level parameters at time $t$. The matrix $\mathbf{T}$ can be interpreted as the local covariance of the observational level parameters at each time $t$. 

Finally, the correlation matrix, $\boldsymbol{\Sigma}(\phi)$, represents the temporal correlation of each observational level parameter over time. Due to the properties of the separable covariance, each of the observational level parameters has the same temporal structure dictated by the form of $\boldsymbol{\Sigma}(\phi)$. The form of $\boldsymbol{\Sigma}(\phi)$ is general such that any standard temporal correlation function may be specified (e.g., exponential or first order autoregressive). The parameter $\phi$ acts as a temporal tuning parameter describing the strength of correlation across time and can be interpreted within the context of each specific temporal structure. 


\subsection{Specifying Hyperprior Distributions}
\label{sec:hypers}

In order to complete the model specification, we define hyperprior distributions for the introduced parameters such that
\begin{equation} \notag 
\boldsymbol{\delta} \stackrel{\text{}}{\sim} \text{MVN}(\boldsymbol{\mu}_{\delta},\boldsymbol{\Omega}_{\delta}), \quad \mathbf{T} \sim \text{Inverse-Wishart}(\xi,\boldsymbol{\Psi}),\quad \phi \sim \text{Uniform}(\text{a}_{\phi},\text{b}_{\phi}).
\end{equation}
The choice of the entries in $\boldsymbol{\mu}_{\delta}$ can be informative or non-informative depending on the context and user. It is important to judiciously consider the entries of $\boldsymbol{\Omega}_{\delta}$. For simplification, we detail a situation where $\boldsymbol{\Omega}_{\delta}$ is diagonal, $\boldsymbol{\Omega}_{\delta}=\text{Diag}(1000,1000,\upsilon_1,\ldots,\upsilon_q)$. The entries in $\boldsymbol{\Omega}_{\delta}$ that are of importance are those that correspond to $\log(\alpha_{tk})$, since the large variances for $\mu_t$ and $\log(\tau_t)$ induce approximately flat priors. More care is needed in specifying $\upsilon_1,\ldots,\upsilon_{\nu}$. These hyperprior variances are chosen for purposes of regularization, in order to encourage $\log(\alpha_{tk})$ to be in a realistic range. Regularization is a common use of Bayesian priors \citep{gelman2013philosophy}. In particular, these variances are chosen so that $\alpha_{tk}$ do not become larger than $\alpha_k^*$, such that $[\alpha_k^* : \exp\{-\alpha_{k}^* z_k\}=0.5]$ with $z_{k} = \stackrel[i,j]{}{\min}\{z_{ijk}\}$. This condition comes from Equation \ref{eq:weights}, where we isolate each dissimilarity metric individually.

For our prior on $\mathbf{T}$, we use an inverse-Wishart distribution with degrees of freedom $\xi=(q+2)+1$ and scale matrix, $\boldsymbol{\Psi}=\mathbf{I}_{q+2}$. This prior is appealing since it induces marginally uniform priors on the correlations of $\mathbf{T}$ and allows for the diagonals to be weakly informative \citep{gelman2014bayesian}. Finally, we specify the hyperprior distribution for the temporal tuning parameter $\phi$ for correlation structures with one parameter. The bounds for $\phi$ cannot be specified arbitrarily since it is important to account for the magnitude of time elapsed. We specify the following conditions for finding the bounds, $[a_{\phi}:[\boldsymbol{\Sigma}(a_{\phi})]_{t,t'}=0.95 , |x_t-x_{t'}| = x_{\text{max}}]$ and $[b_{\phi}:[\boldsymbol{\Sigma}(b_{\phi})]_{t,t'}=0.01 , |x_t-x_{t'}| = x_{\text{min}}]$, where $x_{\text{min}}$ and $x_{\text{max}}$ are the minimum and maximum temporal differences between visits. These conditions state that the lower bound of $\phi$ is small enough so that the greatest length of time between time points can yield a correlation of 0.95 and the upper bound is set so that the shortest length of time between time points can reach 0.01.  These conditions were specified so that $\phi$ can dictate a temporal correlation close to independence ($\phi \rightarrow b_{\phi} $) or strong correlation ($\phi \rightarrow a_{\phi}$), resulting in a weakly informative prior distribution.


\subsection{Prediction}
\label{sec:pred}

Once posterior samples have been obtained, prediction is often a priority. In particular, obtaining samples from the posterior predictive distribution (PPD) is of interest, $f\left(\mathbf{Y}_{\nu+1}|\mathbf{Y}\right)$,

\noindent where $\mathbf{Y}_{t} = (Y_{1t},\ldots,Y_{n_t t})^T$ and $\mathbf{Y} = (\mathbf{Y}_1,\ldots,\mathbf{Y}_{\nu})^T$. We express the PPD as an integral $\int_{\Omega} f\left(\mathbf{Y}_{\nu+1}|\boldsymbol{\Omega},\mathbf{Y}\right) f\left(\boldsymbol{\Omega}|\mathbf{Y}\right) d\boldsymbol{\Omega}$ and then further partition the integral,
\begin{equation} \label{eq:prediction}
\int_{\Omega} \underbrace{f\left(\mathbf{Y}_{\nu+1}|g^{-1}\left(\boldsymbol{\varphi}_{\nu+1}\right),\boldsymbol{\zeta}\right)}_{1} \underbrace{f\left(\boldsymbol{\varphi}_{\nu+1}|\boldsymbol{\theta}_{\cdot \nu+1}\right)}_{2} \underbrace{f\left(\boldsymbol{\theta}_{\cdot \nu+1}|\boldsymbol{\theta},\boldsymbol{\delta},\mathbf{T},\phi\right)}_{3} \\
\underbrace{f\left(\boldsymbol{\zeta},\boldsymbol{\theta},\boldsymbol{\delta},\mathbf{T},\phi|\mathbf{Y}\right)}_{4} d\boldsymbol{\Omega},
\end{equation}
where $\boldsymbol{\Omega}=(\boldsymbol{\varphi}_{\nu+1},\boldsymbol{\theta}_{\cdot \nu + 1},\boldsymbol{\zeta}, \boldsymbol{\theta},\boldsymbol{\delta},\mathbf{T},\boldsymbol{\phi})$.
The convenient form of Equation \ref{eq:prediction} is a function of four known densities that are defined as a consequence of the methodology introduced in Section \ref{sec:methods}. As such, the PPD can be obtained through composition sampling \citep{tanner1996tools}. This theory is presented with one future time point, but is easily generalized to multiple. Full prediction theory details are presented in the online supplementary materials.


\section{ANALYSIS OF VISUAL FIELD DATA}
\label{sec:da}


\subsection{Study Data}
\label{sec:data}

In this study, we source data from the VPSG and the Lions Eye Institute trial registry, Perth, Western Australia. The dataset contains 1,448 VFs from 194 distinct eyes (98 patients in total). Three of the eyes had no clinical assessment of progression and are discarded, yielding 191 series of VFs for analysis. All of the subjects have some form of POAG. The mean follow-up time for participants is 934 days (2.5 years) with an average of 7.4 tests per subject. The progression status of each eye was determined by a group of expert clinicians. Although there is no consensus gold standard for diagnosing progression, there is precedent for treating clinician expertise as a gold standard when introducing new analytic models \citep{betz2013spatial,warren2016statistical}. Every VF series is diagnosed as progressing based on the clinical judgment of two independent clinicians. In the case that the two clinicians disagree, a third clinician is consulted (occurred for only 13 VF series). In our study we have 141 (74\%) stable and 50 (26\%) progressing patient eyes. For a detailed description of the data, please refer to \cite{betz2013spatial}. 


\subsection{Accounting for Zero-Truncation}
\label{sec:zero}

We apply our newly developed methodology to a longitudinal series of VFs. From Section \ref{sec:vf}, we know that VF testing machinery does not allow observations below 0 dB, and therefore any zero measurement represents a potentially censored observation. This motivates the use of a Tobit model \citep{tobit1958estimation}, in which there is precedent in glaucoma progression research \citep{betz2013spatial,bryan2015global}. 

Define the observed DLS, $Y_{it}$, at VF location $i$ with $i=1,\ldots,52$ and visit $t$ with $t=1,\ldots,\nu$. There are 52 VF locations excluding the blind spot and $\nu$ is the number of visits a patient accrues and is patient specific. To induce the Tobit model, define $g(\vartheta_{it})=\varphi_{it}$ with identity link and specify,
\begin{equation} \notag 
f(Y_{it};\vartheta_{it})=P(Y_{it}=x|\vartheta_{it})=1(x=0)1(\vartheta_{it} \leq 0)+1(x=\vartheta_{it})1(\vartheta_{it} > 0),\hspace{0.2in} x\geq0.
\end{equation}
This specification induces the standard Tobit model, $Y_{it}=\max\left\{0,\vartheta_{it}\right\}$, where $\vartheta_{it}$ is an underlying normally distributed latent process.


\subsection{Creating a Dissimilarity Metric}
\label{sec:dm}

\begin{figure}[t]
\begin{center}
\includegraphics[scale=0.90]{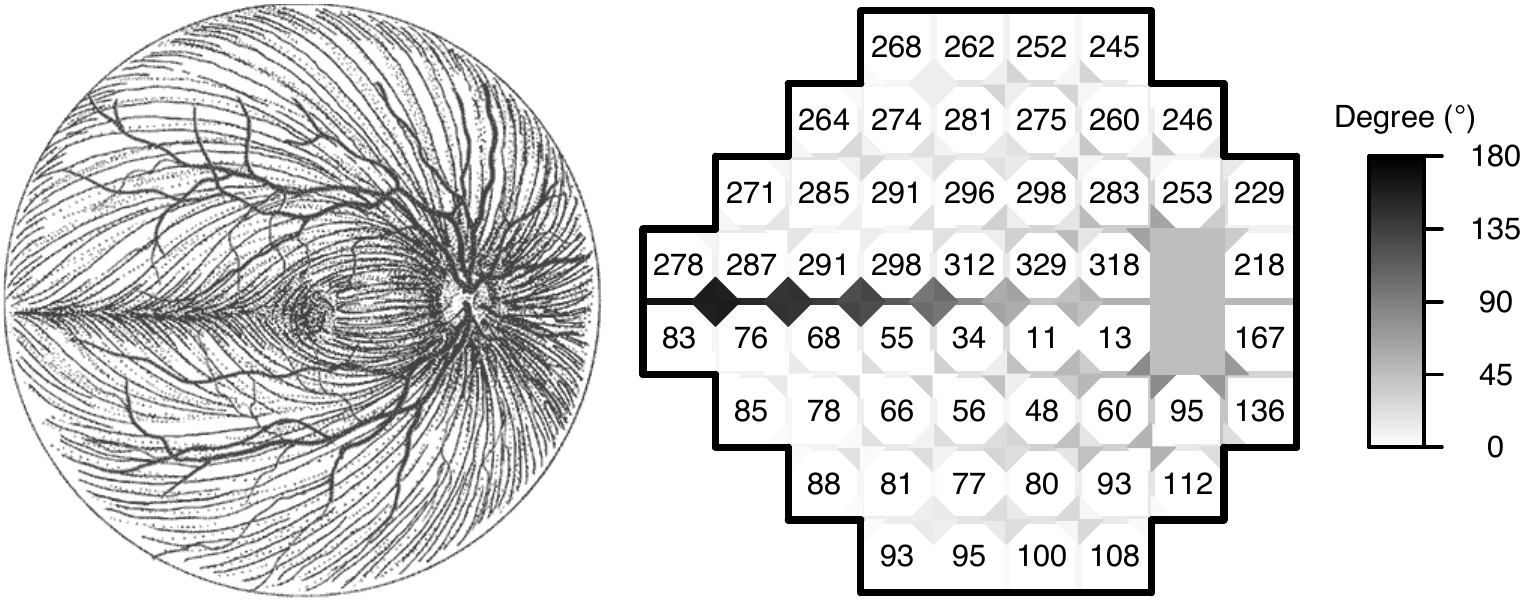}
\caption{Demonstrating the Garway-Heath dissimilarity metric across the VF (right). The angles are defined as the degree at which the underlying nerve fiber enters the optic disc for each location. The edges and corners are shaded to represent the distance between bordering locations on the VF, where darker shading represents a larger distance. The RNFL is displayed to demonstrate the similarity of the pattern that appears on the VF by using the Garway-Heath angles (left).\label{fig:dm}}
\end{center}
\end{figure}

We specify a dissimilarity metric based on the Garway-Heath angles defined in Section \ref{sec:vf}, since we know that correspondence between VF test locations and their underlying nerve fibers is important in determining local neighborhood structure. Figure \ref{fig:dm} displays the dissimilarity metric. It shows that two locations on the VF may be neighbors, but can still be dissimilar in terms of the Garway-Heath angles. Interestingly, a pattern emerges across the VF that mirrors the RNFL (left in Figure \ref{fig:dm}). In particular, the locations that are separated by the superior and inferior regions are separated by nearly 180$^\circ$ and the flow of spatial dependency emulates the nerve fibers. 

Formally, we define $z_{i}$ as the Garway-Heath angle for location $i$. Then, the dissimilarity metric between locations $i$ and $j$ is $z_{ij}=||z_{i}-z_{j}||$. We use the following distance metric, $||x-y||=\text{min}\{|x-y|,360-\text{max}\{x,y\}+\text{min}\{x,y\}\}$. This metric calculates the minimum difference in Garway-Heath angles on the arc of the circular optic disc. We define our dissimilarity parameter at time $t$ as $\alpha_{tGH}$, since we only use a single dissimilarity metric. We allow the event $1(i\sim j)$ to include both edges and corners (i.e., a queen specification).


\subsection{Model Estimation}
\label{sec:estimation}

To finalize the model, we specify the temporal correlation structure and hyperparameters. We define an exponential correlation structure, such that $[\boldsymbol{\Sigma}(\phi)]_{t,t'} = \exp\{-\phi |x_t - x_{t'}|\}$ where $x_t$ is the number of days at visit $t$ after the initial visit, with $x_1 = 0$ for each patient. Based on the criterion in Section \ref{sec:hypers} for the Garway-Heath dissimilarity metric, we specify $\upsilon_1=1$. Then we set $\boldsymbol{\Omega}_{\delta} =\text{Diag}(1000,1000,1)$ and $\boldsymbol{\mu}_{\delta}=(3,0,0)$. For complete details on the implementation of the model, see Section 1 of the online supplementary materials.


\subsection{Diagnosing Glaucoma Progression}
\label{sec:diag}

\subsubsection{Establishing a Novel Diagnostic Metric}

\begin{figure}[ht]
\begin{center}
\includegraphics[scale=0.80]{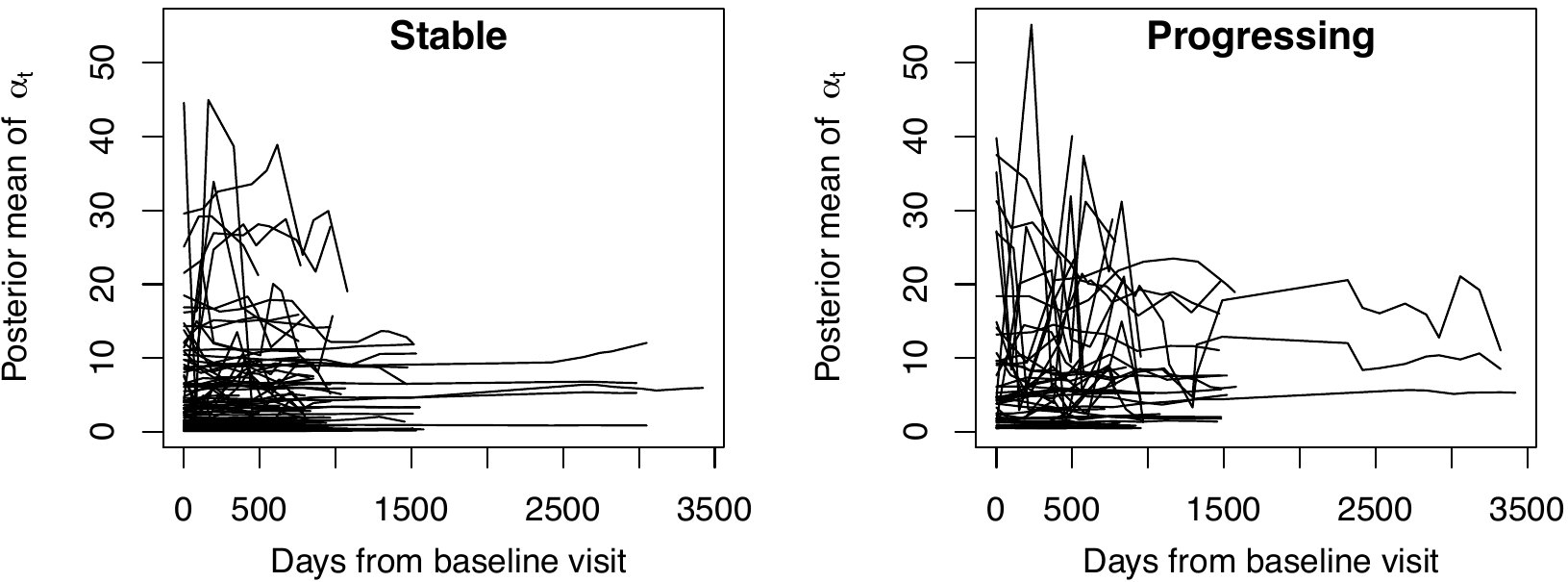}\\
\caption{ Posterior mean estimates of $\alpha_{tGH}$ plotted across time for all patients, subset by progression status defined at the end of the study. \label{fig:alphaovertime}}
\end{center}
\end{figure}

The methodology presented in Section \ref{sec:methods} provides a novel framework for studying glaucoma progression. Since progression is characterized by worsening disease severity over time, we propose using a function of $\alpha_{tGH}$ that can quantify this variation. We suggest using the posterior mean of the coefficient of variation (CV) of $\alpha_{tGH}$ for $t=1,\ldots,\nu$. The CV is the ratio of the standard deviation and mean and is an ideal summary metric, because it accounts for the variability in a parameter while standardizing by its mean, allowing it to be comparable over populations. This metric is novel clinically as $\alpha_{tGH}$ does not describe mean trend in DLS over time, but rather is representative of optic disc damage and changes in the spatial covariance structure across time.  We refer to our metric based on the spatiotemporal (ST) model as ``ST CV''. In Figure \ref{fig:alphaovertime}, we display posterior mean estimates of $\alpha_{tGH}$ over time for all patients, subset by progression status.  The figure suggests that there is more variability in the estimates across time for the progressing patients generally, and that there is not a clear trend in the estimates across time for either group.  This further motivates a metric like CV to quantify general variability across time for diagnostic purposes.

In order to assess the novelty of our metric, we compare it to two metrics that aim to be representative of a class of standard VF trend-based methods. Trend-based VF diagnostic techniques can be grouped in two categories, global and point-wise, and study associations between VF changes and progression \citep{vianna2015detect}. We define the global metric ``Mean CV" as the CV of the VF wide mean of DLS at each visit. If the patient is stable, the means should be similar at each visit and the CV of the metric should be near zero. On the other hand, a progressing patient will have varying means at each visit and the CV should increase as visits accrue. We also compare our model to a commonly used point-wise linear regression (PLR) method.  The progression metric is defined as the minimum p-value for the slope parameter from separately run simple linear regression analyses across all VF locations, where observed DLS is the dependent variable and time from first visit is the independent variable \citep{smith1996}. Using both ``Mean CV'' and PLR as comparators is far-reaching, as global methods are generally more robust, while point-wise methods identify local changes in visual ability.

To assess the diagnostic capability of our metric, we construct logistic regression models regressing various metrics on the clinical assessment of progression. We compare our method to ``Mean CV", PLR, and also the posterior mean CV of $\alpha_{tGH}$ from the \cite{lee2011boundary} model, where $\alpha_{tGH}$ is estimated independently at each visit. We refer to this metric as ``Space CV". To obtain ``Space CV", we apply the \cite{lee2011boundary} methodology with a Tobit likelihood. Thus, the appreciable differences between the two models that produce ``Space CV'' and ``ST CV'' are the definition of the weights, cross-covariance, and temporal correlation structure for the observational level parameters.

\begin{table}[t]
\centering
\caption{Assessing the diagnostic capability of VF metrics. Each metric is regressed against the clinical assessment of progression using a logistic regression model with the slope estimates being displayed ($^*$significance level of 0.05). ``Mean CV'' is the CV of the VF wide mean of DLS at each visit, PLR is the minimum p-value of the location specific regression slopes, and ``Space CV'' and ``ST CV'' represent the mean posterior CV of $\alpha_{tGH}$ from the \cite{lee2011boundary} and our spatiotemporal models, respectively. Each of the metrics are standardized.\label{table:metrics}}
\begin{tabular}{lrrrl}
  \hline
Metrics & Estimate & Std. Error & z value & Pr($>$$|$z$|$) \\ \hline
Mean CV  &  {0.40} & {0.16} & {2.55} & {0.011$^*$} \\
{PLR} &  {-0.59} & {0.25} &  {-2.37} & {0.018$^*$} \\
Space CV & {-0.07} & {0.17} & {-0.41} & {0.680}     \\ 
ST CV    &  {0.39} & {0.16} & {2.44} & {0.015$^*$} \\ \hline
\end{tabular}
\end{table}

In Table \ref{table:metrics} we present the results from the logistic regression analyses using a significance level of 0.05. Each predictor is standardized in order to facilitate comparisons of the different metrics. We can see that as expected ``Mean CV'' and PLR are significantly associated with glaucoma progression, with p-values of 0.011 and 0.018, respectively. For ``Mean CV'', the estimated slope coefficient of 0.40 indicates that as a patient's ``Mean CV" increases, their risk of progression increases. For PLR, a smaller minimum p-value suggests an increased risk of progression (estimated slope coefficient of -0.59). Based on the glaucoma literature, we would expect trend-based methods such as ``Mean CV" and PLR to have good discriminatory capability between stable and progressing eyes. It was less clear for our new metrics dependent on $\alpha_{tGH}$. We see that ``Space CV'' is not significantly associated with glaucoma progression with a p-value of 0.680, while ``ST CV'' is significantly associated with a p-value of 0.015. This result is encouraging, yet surprising, since the models are similar and illuminates their differences in the VF data setting. However, for this newly defined metric to be impactful, we must verify that it is explaining a novel pathway in glaucoma progression, independent of existing metrics. 

In order to assess whether ``ST CV'' provides novel diagnostic capabilities, we explore the correlation between the metrics. In Figure 1 of the online supplementary materials, we show pairwise correlation plots. We also present Pearson correlation estimates $(\rho)$ and p-values from the hypothesis test: $H_0: \rho = 0, \quad H_1: \rho \neq 0$. ``ST CV'' is uncorrelated with both ``Mean CV'' and PLR with estimated correlations of 0.06 and 0.11, respectively, and large p-values. This result has important implications; indicating that in addition to being highly predictive of progression, ``ST CV'' is uncorrelated with the standard trend-based metrics.  This suggests that ``ST CV'' and the trend-based metrics can be used in conjunction in order to diagnose progression.  Finally, the estimated correlation between ``Space CV'' and ``ST CV'' is 0.48 and the p-value is highly significant ($<0.001$), which is not surprising since ``Space CV'' and ``ST CV'' are estimating the same quantity. It is, however, interesting that these two metrics have such different associations with glaucoma progression (see Table \ref{table:metrics}). When ``Space CV'' was calculated with continuous weights this association did not change (p-value: 0.326, not included in Table \ref{table:metrics}), indicating the importance of temporal correlation and cross-covariance for properly modeling VF data. Presumably due to our enhanced methodology, ``ST CV'' is more precisely estimating the CV of $\alpha_{tGH}$ by smoothing the $\alpha_{tGH}$ and eliminating temporal noise and cross-covariance dependencies. To formalize this hypothesis a simulation study is designed in Section \ref{sec:sim}. 


\subsubsection{Extension to the Clinical Setting}
\label{sec:clinical}

Having demonstrated the novelty of using variability in the boundary detection parameter as a progression diagnostic, we now establish its clinical utility. We combine the two trend-based metrics into a composite model that represents an ideal synthesis of global and local methods. This metric includes ``Mean CV'' and PLR and their interaction. We then compare the changes in operating characteristics after separately adding ``Space CV'' and ``ST CV'' to this composite model, including the main effect and pairwise interactions. These results are presented in Table \ref{tab:op},
\begin{table}[t]
\centering
\caption{Operating characteristics for diagnostic metrics. P-values assess statistical significance of improvement over the combined trend-based metric (Mean CV \& PLR) for, 1) additionally included regression parameters (using a nested likelihood ratio test), 2) AUC, and 3) pAUC. \label{tab:op}}
\begin{tabular}{lrrrccc}
  \hline
 & AIC & AUC & pAUC & p-value$_1$ & p-value$_2$ & p-value$_3$ \\ 
  \hline
  Mean CV \& PLR & 209.81 & 0.68 & 0.21 & \multicolumn{1}{c}{$-$} & \multicolumn{1}{c}{$-$} & \multicolumn{1}{c}{$-$} \\ 
  Mean CV \& PLR + Space CV & 213.46 & 0.70 & 0.22 & 0.503 & 0.196 & 0.410 \\ 
  Mean CV \& PLR + ST CV & 204.46 & 0.74 & 0.29 & 0.010 & 0.042 & 0.059 \\ 
   \hline
\end{tabular}
\end{table}
where summary statistics include the Akaike information criterion (AIC), area under the receiver operating characteristic (ROC) curve (AUC), and the partial AUC (pAUC). Here, pAUC is limited to the clinically significant region of specificity of 85-100\% \citep{zhu2014detecting}.

\begin{figure}[!ht]
\begin{center}
\includegraphics[scale=1]{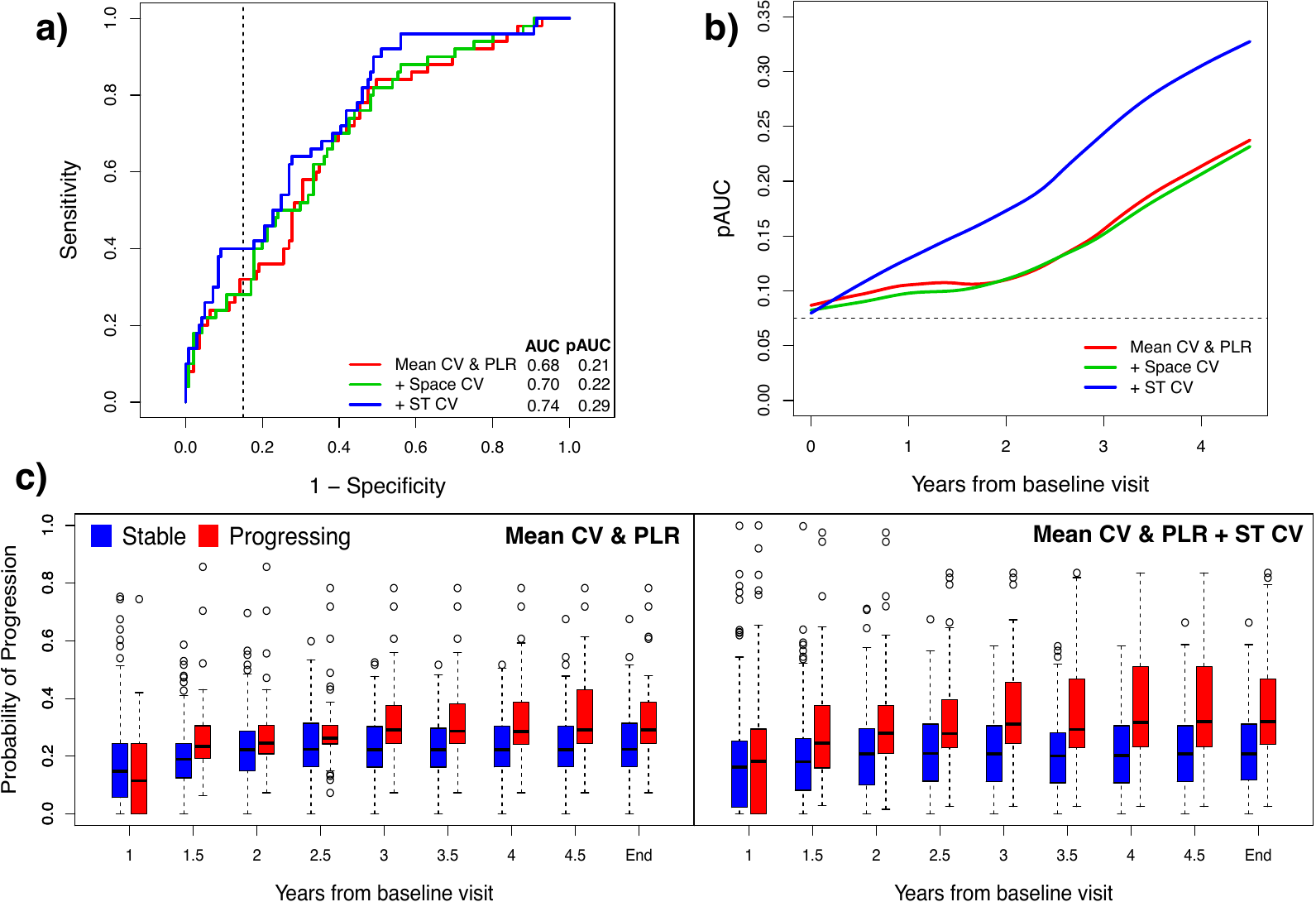}\\
\caption{Demonstrating the performance of diagnostic metrics: \textbf{a)} ROC curves for the combined trend-based metric (Mean CV \& PLR), including the addition of ``Space CV" and ``ST CV" (clinically meaningful region to the left of dashed line), \textbf{b)} pAUC in the initial years from baseline visit (estimates are presented as smooth LOESS curves and the dashed line indicates no discrimination), \textbf{c)} boxplots of the predicted probabilities of progression presented by disease status over time as defined at the end of the study period.\label{fig:roc}}
\end{center}
\end{figure}

The results indicate the clinical importance of ``ST CV'', producing optimal values of AIC, AUC, and pAUC. Furthermore, adding ``ST CV'' produces statistically significant improvements over the composite trend-based model in terms of the additional predictors, and both AUC and pAUC (although at the $\alpha = 0.10$ level for pAUC). Meanwhile, there are no improvements in the operating characteristics when adding ``Space CV'', which is also confirmed when investigating the ROC curves (Figure \ref{fig:roc}a). In Figure \ref{fig:roc}a, it is clear that the inclusion of ``ST CV'' increases the discriminatory ability of the composite trend-based model, in particular within the clinically significant region, left of the dashed line.

In addition to operating characteristics calculated at the end of the study, we are also interested in the diagnostic performance of each metric in earlier stages of the disease, where accurate discrimination is more important for preserving a patient's visual ability. In Figure \ref{fig:roc}b, a smoothed LOESS trajectory of pAUC is presented for each of the models in Table \ref{tab:op} for the first 4.5 years of follow-up. This analysis reflects a true clinical setting where data are analyzed as they are collected for a patient over time. We calculate ``Mean CV", PLR, ``Space CV", and ``ST CV" for each patient during each visit. Then, using the fitted logistic regression models from the end of the study, we calculate predicted probabilities of VF progression. Finally, these probabilities are converted into binary disease progression diagnoses based on probability cutoffs obtained from the full study data. We forgo the commonly used Youden's index (which maximizes sensitivity and specificity), for a criteria that forces specificity to be in the clinically significant range, while maximizing sensitivity.

These results indicate that the improvement in operating characteristics attributed to ``ST CV'' begins early during the follow-up period, in particular for pAUC. This result establishes that ``ST CV'' is clinically useful, and further enforces the importance of the introduced methodology as ``Space CV'' is incapable of improving operating characteristics during the same time period. In addition, Figure \ref{fig:roc}c presents boxplots of the predicted probabilities at half-year intervals, by progression status as defined at the end of the study. In Figure \ref{fig:roc}c, there is a clear separation between the progressing and non-progressing patients' predicted probabilities of progression beginning around 1.5 years after baseline visit. This separation is more pronounced with the addition of ``ST CV''.


\subsection{Sensitivity Analyses}
\label{sec:sa}

In the online supplementary materials, we present a number of sensitivity analyses and an additional simulation study to explore various modeling assumptions, including hyperparameter choices, the correlation structure, the bounds of $\phi$, and misspecification of $\rho$ and the dissimilarity metric. Overall, we find that the results are robust to these assumptions.  For full details on these analyses, please see the online supplementary materials.


\section{MODEL PERFORMANCE SIMULATION}
\label{sec:sim}

A simulation study is designed to assess the performance of the proposed model in the presence of temporal dependence and cross-covariance. We focus on the estimation of CV of $\alpha_{tGH}$, since we propose using this posterior distribution in diagnosing progression.

In order to understand the gains of using our methodology in the presence of temporal correlation and cross-covariance dependence, we design a simulation study comparing the spatial model of \cite{lee2011boundary}, referred to as Space, and our spatiotemporal method (ST). We simulate data based on a set of known hyperparameters and then estimate the posterior mean of the CV of $\alpha_{tGH}$ using both models, comparing the estimates to the known truth using bias, mean squared error (MSE), and empirical coverage (EC). The simulation is designed to explore model performance in a typical patient from our study data. As such, we fix the true hyperparameters in the simulation study to the posterior means obtained from Section \ref{sec:da} for an average patient. Here an average patient is one whose posterior mean estimates are average amongst all patients. The true hyperparameters used in the simulation study are as follows,
$$\boldsymbol{\delta} = \left[\begin{array}{r}
2.446 \\
0.070 \\
0.974 
\end{array} \right], \quad \mathbf{T}=\left[\begin{array}{rrr}
 0.820 & 0.004 &-0.028 \\
 0.004 & 0.380 &-0.191 \\
-0.028 &-0.191 & 0.840 
\end{array} \right], \quad
\phi = 0.163.$$

Simulation settings are developed to incrementally understand the impact of the cross-covariance, $\mathbf{T}$, and temporal correlation, $\boldsymbol{\Sigma}(\phi)$. To facilitate this analysis, Equation \ref{eq:separable} is utilized. Define $\mathbf{T}_{\text{Diag}} = \text{Diag}(\mathbf{T})$, such that $\mathbf{T}_{\text{Diag}}$ has zeros on the off-diagonal and $\phi_I = 100$ (note that a large $\phi$ for the exponential correlation implies temporal independence). The simulation settings are as follows with the data generating covariance given in parentheses, A: no temporal correlation, no cross-covariance $\left(\boldsymbol{\Sigma}\left(\phi_I\right) \otimes \mathbf{T}_{\text{Diag}}\right)$, B: no temporal correlation, cross-covariance $\left(\boldsymbol{\Sigma}\left(\phi_I\right) \otimes \mathbf{T}\right)$, C: temporal correlation, no cross-covariance $\left(\boldsymbol{\Sigma}\left(\phi\right) \otimes \mathbf{T}_{\text{Diag}}\right)$, D: temporal correlation, cross-covariance $\left(\boldsymbol{\Sigma}\left(\phi\right) \otimes \mathbf{T}\right)$. Finally, we must specify the days and number of VF visits. To obtain the visit days, we sampled from a Poisson distribution with rate parameter equal to the average difference in days between VF visits (rate = 117.25 days). We present the simulation at the median (7) and maximum (21) number of visits in our study data. 


For each simulation setting, we generate 100 values of $\boldsymbol{\theta}$ and then use each $\boldsymbol{\theta}$ to simulate 10 datasets. This yields 1,000 simulated datasets for each of the simulation settings. In Table \ref{table:sim}, the bias, MSE, and EC are presented for the two models across all settings and at the median and maximum number of visits. We begin by noting the average (across all settings) simulation standard errors of the bias (0.223, 0.063), MSE (0.039, 0.028) and EC (0.141, 0.013) for the space and spatiotemporal models, respectively. In general the spatiotemporal model has smaller standard errors than the space model, indicating less variability in estimation across the simulated datasets.

\begin{table}[t]
\centering
\caption{Results from simulation study estimating the posterior mean of the CV of $\alpha_{tGH}$ in setting A: no temporal correlation, no cross-covariance, B: no temporal correlation, cross-covariance, C: temporal correlation, no cross-covariance, and D: temporal correlation, cross-covariance. Each setting is also implemented for the median (7) and maximum (21) number of VF visits. Each reported estimate is based on 1,000 simulated datasets.\label{table:sim}}
\begin{tabular}{llrrrcrrr}
  \hline
& & \multicolumn{7}{c}{\# Visits} \\ 
& & \multicolumn{3}{c}{7 (Median)} & & \multicolumn{3}{c}{21 (Maximum)} \\ \cline{3-9}
Setting & Model & \multicolumn{1}{c}{Bias} & \multicolumn{1}{c}{MSE} & \multicolumn{1}{c}{EC} & & \multicolumn{1}{c}{Bias} & \multicolumn{1}{c}{MSE} & \multicolumn{1}{c}{EC} \\ \hline
A & ST & 0.032 & 0.107 & 0.97 & & 0.023 & 0.084 & 0.95 \\ 
  & Space  & 0.102 & 0.111 & 0.87 & & 0.174 & 0.098 & 0.81 \\ \hline
B & ST & 0.047 & 0.125 & 0.96 & & 0.034 & 0.101 & 0.95 \\ 
  & Space  & 0.119 & 0.120 & 0.80 & & 0.250 & 0.176 & 0.68 \\ \hline
C & ST &-0.113 & 0.088 & 0.97 & & 0.002 & 0.037 & 0.98 \\ 
  & Space  &-0.306 & 0.172 & 0.51 & &-0.182 & 0.088 & 0.60 \\ \hline
D & ST &-0.103 & 0.085 & 0.98 & & 0.015 & 0.060 & 0.98 \\ 
  & Space  &-0.299 & 0.172 & 0.51 & &-0.169 & 0.087 & 0.59 \\ \hline
\end{tabular}
\end{table}

The bias results suggest that our spatiotemporal model produces an estimator of CV with bias closer to zero than the spatial model on average across all settings. In the settings with no temporal correlation (i.e., A and B) the MSE is quite similar between the models, except for Setting B where the maximum number of visits is considered. In this setting, the MSE is lower for the spatiotemporal model, revealing the importance of the cross-covariance for properly estimating the posterior mean of the CV of $\alpha_{tGH}$. In all other settings, the estimated MSEs for the spatiotemporal model are smaller than for the spatial model, with the largest differences seen in Settings C and D for the median number of visits.

The EC is the proportion of the time that the estimated Bayesian credible intervals contain the true CV value. We define the nominal coverage as 95\%. The EC results favor the spatiotemporal model across all settings. The spatial model has an EC of 0.87 and 0.81 in setting A for the median and maximum settings, respectively. This is the most ideal setting for the spatial model, because it was the setting in which the model was introduced (although within the context of disease mapping). The EC deteriorates in the spatial model as cross-covariance and temporal structure are introduced in the simulated data, falling as low as 0.51, while the spatiotemporal model performs consistently.


\section{DISCUSSION}
\label{sec:disc}

In this paper, we proposed a modeling framework for incorporating local neighborhood structure into complex spatiotemporal areal data. Based on this framework we developed an innovative and highly predictive diagnostic of glaucoma progression that outperformed the spatial-only model. Although motivated by VF data, the methodology was introduced in a general manner that permits the model to be applied in broad areal data settings, including disease mapping and more generally GLMM. The methodology is built upon theory in boundary detection literature, utilizing a Bayesian hierarchical modeling framework for inference. We extended the spatial-only method introduced by \cite{lee2011boundary}, that elegantly introduced a dissimilarity metric in a parsimonious framework. Our method allows for the local neighborhood structure to adapt over time as a function of changing dissimilarity metric parameters. Furthermore, the temporal correlation and cross-covariance are accounted for, eliminating known sources of excess variability. This parsimonious method induces non-standard local spatial surfaces in areal data that are capable of mirroring complex processes, such as the surface of the VF. 

We have shown (Figure \ref{fig:dm}) that the spatiotemporal covariance structure specified in our methodology successfully induces local neighborhood structure across the VF. The novelty of employing the Garway-Heath angles in the form of a dissimilarity metric provides a connection between the VF and the underlying optic disc, resulting in a neighborhood structure on the VF that is representative of the RNFL. Other methods incorporated the Garway-Heath angles into statistical models, either collapsed into regions of anatomical interest \citep{betz2013spatial,warren2016statistical} or by VF location  \citep{zhu2014detecting}, but none allowed the effect to change dynamically. The dissimilarity metric parameter, $\alpha_{tGH}$, dictates the VF spatial surface at each visit, allowing the neighborhood structure to adapt alongside changes in DLS.  

The results from applying our method to VF data (Section \ref{sec:da}) demonstrated the added benefit of using our methodology, in both the clinical and statistical frameworks. We defined the diagnostic metric ``ST CV" as the posterior mean of the CV of $\alpha_{tGH}$, and showed that it is a significant predictor of clinically determined glaucoma progression while being uncorrelated with standard VF trend-based metrics (``Mean CV", PLR). Since ``ST CV'' is independent of ``Mean CV" and PLR, and each one is an effective predictor, there is clinical utility to combining their diagnostic capability (i.e.,``ST CV" is not meant to compete with ``Mean CV" and PLR but should be used together). The ``Space CV" metric was not significantly associated with progression.

The significant association observed between ``ST CV" and glaucoma progression was statistically note-worthy, but taken in isolation had limited clinical implications. We showed the clinical utility of ``ST CV" in Section \ref{sec:clinical}, where the operating characteristics of the trend-based methods, both at the end of the study and in the early period, are improved with the addition of ``ST CV'' (Table \ref{tab:op}, Figure \ref{fig:roc}). This fundamental finding shows that ``ST CV'' constitutes an alternate pathway for studying glaucoma progression using VF data. This pathway is facilitated by the dissimilarity metric framework introduced by the proposed methodology and consequently ``ST CV" has a novel interpretation amongst glaucoma progression diagnostic metrics. In particular, ``ST CV'' quantifies damage to the optic disc over time as a function of stability on the spatial structure of the VF. It provides a method for discussing underlying damage to the optic disc using VF data. 

Our simulation study indicated that the presence of temporal correlation and cross-covariance dependence impacted estimation of the posterior mean of the CV of $\alpha_{tGH}$. The results (Table \ref{table:sim}) illustrated superior bias, MSE, and EC for our proposed spatiotemporal model over the spatial model of \cite{lee2011boundary}. These results imply that our method provides a framework for more precisely estimating the CV of $\alpha_{tGH}$ by smoothing the $\alpha_{tGH}$ and eliminating temporal noise and cross-covariance dependencies. These results are consistent with studies that have shown when temporal correlation is ignored, estimators are often biased and variances are poorly estimated \citep{west1996bayesian}.

In our implementation of the model for glaucoma data, we are not particularly interested in the specific location of the spatial boundaries across the VF, unlike traditional boundary detection applications. In isolation we do not believe those findings would be predictive of glaucoma progression, but would only inform about the current level of vision loss for a patient. Our results suggest that changes in these boundaries across time represent an informative, innovative, and unique metric for diagnosing glaucoma progression. Using the dissimilarity metric gives us the ability to directly quantify the boundary changes across time and therefore, to make clinical progression determinations for a patient. 

The need for spatiotemporal boundary detection with a dissimilarity metric is also driven by the unique features of our VF dataset. Using a more basic CAR structure with fixed neighborhood adjacencies would result in ignoring the fact that these neighborhood definitions are potentially changing across time due to disease progression. It would also require us to fix the adjacencies \textit{a priori}. On the other hand, using a point-referenced geostatistical model with a spatiotemporal covariance function would result in ignoring the inherent discreteness of the spatial domain, a grid overlaying the VF.  Spatiotemporal boundary detection with a dissimilarity metric represents an ideal blend between the two methods by allowing for discreteness in the spatial domain and flexibility in defining neighbors through use of a continuous covariate such as the Garway-Heath angles.

Finally, this work opens up numerous avenues for future statistical research. If data are available, covariates can be incorporated in the observational model naturally via the link function, $g(\vartheta_{it}) = x_{it} \boldsymbol{\beta} + \varphi_{it}$. Currently, the specification of the \cite{leroux2000estimation} likelihood includes a $\mu_t$ in the mean structure, thus in order to identify the $\mu_t$ and $\beta_0$ we must either apply a constraint, $\sum_{t=0}^{\nu}\mu_t = 0$, or re-parameterize the likelihood to be a function of only $\tau_t$ and $\boldsymbol{\alpha}_t$. The incorporation of covariates is of particular importance in disease mapping, where standard incidence rates are often mapped over varying risk factors \citep{lee2013locally}. Another natural extension to this approach includes generalizing the separable temporal covariance to allow unique temporal decay parameters. 


\section*{SUPPLEMENTARY MATERIAL}
\label{sec:sm}

The supplementary materials contain details for implementing the MCMC sampler, including derivation of full conditionals, along with prediction theory, additional sensitivity analyses, and figures.

\bibliographystyle{References/Chicago.bst}

\bibliography{References/references.bib}

\end{document}